\documentclass[pra,aps,onecolumn,floatfix,
               showpacs,noshowpacs,
               balancelastpage,notitlepage,
               longbibliography,
               superscriptaddress, final]{revtex4-1}

\usepackage{amsthm}
\usepackage{amsmath}
\usepackage{amssymb}
\usepackage{mathtools}
\usepackage{physics}
\usepackage[table,xcdraw]{xcolor}
\usepackage{graphicx}
\usepackage{grffile}
\usepackage[T1]{fontenc}
\usepackage{csquotes}
\usepackage{longtable}
\usepackage[pdftex, pdftitle={Article}, pdfauthor={Author}]{hyperref}
\usepackage{epsfig}
\usepackage{multirow}
\usepackage{bm}
\usepackage{color}
\usepackage{url}

\usepackage[table,xcdraw]{xcolor}

\graphicspath{{figures/}} 

\newcommand{\beq}{\begin{equation}}
\newcommand{\eeq}{\end{equation}}
\newcommand{\bea}{\begin{eqnarray}}
\newcommand{\eea}{\end{eqnarray}}

\newcommand{\fig}[1]{Fig.~\ref{#1}}

\newcommand{\eF}{\varepsilon_{F}}

\newcommand{\kF}{k_{\textrm{F}}}
\newcommand{\kFi}{k_{\textrm{F}}^{-1}}

\newcommand{\Nreq}{N^{\textrm{(req.)}}}

\newcommand{\updown}{\uparrow\downarrow}

\begin{document}
%
%
\title{Disordered structures in ultracold spin-imbalanced Fermi gas}
\author{Bu\u{g}ra T\"uzemen}
\affiliation{Center for Theoretical Physics, Polish Academy of Sciences, Al. Lotnik\'ow 32/46, 02-668
Warsaw, Poland}
\affiliation{Faculty of Physics, Warsaw University of Technology, Ulica Koszykowa 75, 00-662 Warsaw}
\author{Tomasz Zawi\'slak}
\affiliation{Faculty of Physics, Warsaw University of Technology, Ulica Koszykowa 75, 00-662 Warsaw}
\author{Gabriel Wlaz\l{}owski}
\affiliation{Faculty of Physics, Warsaw University of Technology, Ulica Koszykowa 75, 00-662 Warsaw}
\affiliation{Department of Physics, University of Washington, Seattle, Washington 98195-1560, USA}
\author{Piotr Magierski}
\affiliation{Faculty of Physics, Warsaw University of Technology, Ulica Koszykowa 75, 00-662 Warsaw}
\affiliation{Department of Physics, University of Washington, Seattle, Washington 98195-1560, USA}

\begin{abstract}
We investigate  
properties of spin-imbalanced ultracold Fermi gas in a large range of spin polarizations
at low temperatures. We present results of microscopic calculations based on mean-field and density functional theory approaches,
with no 
symmetry constraints.
At low polarization values we predict the structure of the system as consisting of several spin-polarized droplets. As the polarization increases, the system self-organizes into a disordered structures similar to liquid crystals, and energetically they can compete 
with ordered structures such as grid-like domain walls. At higher polarizations 
the system starts to develop  
regularities
that, in principle, can be called supersolid, where periodic density modulation and pairing correlations coexist. 
The robustness of the results has been checked with respect to temperature effects, dimensionality, and the presence of a trapping potential. Dynamical stability has also been investigated.  
\end{abstract}

\maketitle

\section{Introduction}
The crossover between Bardeen-Cooper-Schrieffer (BCS) type superfluid and Bose-Einstein condensate (BEC) of diatomic pairs has been, in the last years, a subject of considerable theoretical and experimental efforts~\cite{RevModPhys.80.885,RevModPhys.80.1215,Zwerger2011-bj,OHASHI2020103739}.
The research is driven by developments in the field of ultracold Fermi gas. In the case of spin-symmetric gas, the phase diagram as a function of temperature and interaction strength is well-established, apart from the presence of a pseudogap regime which is still the subject of ongoing discussion~\cite{Mueller_2017,universe6110208}.
In contrast, spin-imbalanced systems lack an undisputed phase diagram~\cite{Chevy_2010,Radzihovsky_2010,GUBBELS2013255}. 
There are various proposed phases. For high values of the spin polarization, but still below the critical value above which the system transit into a normal state, a homogeneous gapless superfluidity may emerge~\cite{liu, forbes_2005, stephanov}. An alternative scenario was proposed by Larkin and Ovchinnikov (LO)~\cite{lo} and simultaneously by Fulde and Ferrel (FF)~\cite{ff}. It is described by the spatial modulation of the pairing field (order parameter) $\Delta$ due to the mismatch between the Fermi surfaces of the paired species. 
The FF-type of modulation is described as $\Delta(\textbf{r})\sim|\Delta|e^{i\textbf{q}\cdot\textbf{r}}$ where $\textbf{q}$ is the momentum of the Cooper pair due to the mismatch. It breaks the time-reversal symmetry while preserving the translational symmetry. In contrast, LO-type of modulation breaks the translational symmetry while preserving the time-reversal symmetry and has the form of $\Delta(\textbf{r})\sim|\Delta|\cos(\textbf{q}\cdot\textbf{r})$. This so-called LOFF phase remains one of the long-sought phenomena in ultracold Fermi gases. 
To date, the predictions (both theoretically and experimentally) point out that the LOFF phase occupies a small portion of the phase diagram~\cite{Chevy_2010,Radzihovsky_2010,GUBBELS2013255,bulgac_2006} and it is still an elusive phenomenon.
The presence of the LOFF phase and its stability has been a subject of intensive research. 
In particular, the interplay between LOFF and phase separation as a main competing scenario has been 
discussed~\cite{SHEEHY20071790,PhysRevLett.96.060401}. 

\begin{figure}[b]
	\centering 
    \includegraphics[width=0.98\textwidth]{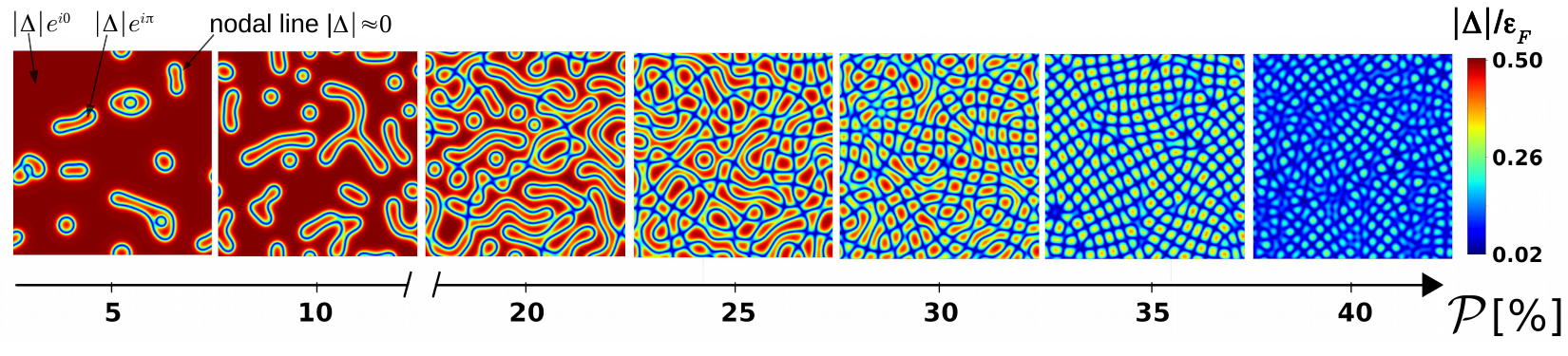}
	\caption{Spatial distribution of absolute value of the order parameter $|\Delta(\textbf{r})|$ 
	for several values of global spin polarization $\mathcal{P}$. They correspond to self-consistent solutions of energies close to the ground state energies.  
The size of the simulation box is $192\kFi\times 192\kFi$, with Fermi momentum $\kF \approx 1$. The coupling constant $g$ was adjusted to mimic strength of the pairing correlation as for the unitary Fermi gas. At nodal lines/regions $\Delta\approx 0$ (blue color) the order parameter changes its sign.}
	\label{fig:pol}
\end{figure}

In the case of the LOFF phase, it is implicitly assumed that the long-range order is set in the system
producing highly symmetric configurations. Their appearance has
consequences for the spectrum of low-energy excitation modes. 
This issue has been raised 
and discussed in Refs.~\cite{PhysRevA.84.023611, PhysRevLett.103.010404} where low-energy effective Hamiltonian has been
proposed assuming
a certain type of ordering or the deviations from regular structures. 
In most of these studies, Ginzburg-Landau model is employed based on its relation to  microscopic 
approach~\cite{PhysRevA.84.023611,PhysRevB.99.220508,BUZDIN1997341,Agterberg_2001}. However, little is known about the possible structures that lacks regularities, in particular whether they can compete with the ordered structures. 
This brings us to the question concerning energetics and stability of 
configuration with broken symmetries.
Although the issue of stability has been 
raised by various authors, it is usually considered in terms of expansions around
the symmetric configuration \cite{PhysRevA.84.023611,PhysRevLett.103.010404,RevModPhys.76.263}, whereas highly nonordered structures have never been studied.
There are two aspects that require to shed light upon. The first one concerns
the energy differences between ordered and highly nonordered structures and their susceptibility
with respect to thermal fluctuations.
The second one concerns the typical time scales and 
the character of the evolution of these structures. Insight into both aspects can be provided by microscopic approach which explicitly takes into account fermionic
degrees of freedom to account for possible shell effects coming from single-quasiparticle states. At the technical level, applying the framework is very challenging as one needs to execute calculations with no symmetry constraints. Nevertheless, such studies are particularly timely 
in the light of experimental efforts to detect LOFF-like structures~\cite{Zwierlein2006,PhysRevLett.97.030401,Mukherjee2017,PhysRevA.92.063616,PhysRevLett.117.235301,PhysRevA.102.033311,PhysRevA.96.023612,Kinnunen_2018}. They also
provide information about the characteristics of low-energy modes that may appear in such systems.

In this paper, we focus on the 
solutions of a spin-imbalanced ultracold Fermi gas as a function of a wide range of global polarization $\mathcal{P} = \left(N_{\uparrow} - N_{\downarrow}\right)/\left(N_{\uparrow} + N_{\downarrow}\right)$, where $N_{\uparrow}$, and $N_{\downarrow}$ are the numbers of spin-up and spin-down particles, respectively. In order to explore the richness of possible configurations, we 
break the translational symmetry of the pairing field $\Delta(\textbf{r})$ by perturbing it with a spatial, complex random noise and subsequently 
perform unconstrained energy minimization.
This symmetry breaking procedure can be thought of as mimicking thermal fluctuations
which occur during the cooling process.

The sample of obtained solution is presented in \fig{fig:pol}. 
For low polarization values (e.g., $\mathcal{P}=5\%$), the randomly perturbed system produces several distinct spin-polarized 
droplets called \textit{ferrons}~\cite{ferron1,ferron2,ferron3,babaev}. 
The increase in the spin-imbalance induces  
elongation of nodal lines, which eventually form more complex disordered structures extended over the whole system. Here, for medium polarizations such as $\mathcal{P}=20-25\%$, the nodal structure of the order
parameter 
resembles rather a liquid crystal.
There is a vast number of different nodal structures which are practically energetically degenerate. 
As the total length of nodal lines becomes comparable with the system size, the number of different ways to arrange the nodal lines decreases; the system starts to develop an order in configuration. For high polarization values (i.e., $\mathcal{P}\gtrsim30\%$), the system arranges itself in different regions where each region shows a unique lattice-like character. For extremely high polarization ($\mathcal{P}\gtrsim 40\%$), one may expect to find a single lattice structure that resembles a supersolid. However, as the polarization increases, the mean distance between the nodal lines becomes comparable to the coherence length $\xi$. This degrades the superfluid phase, and eventually, the system transits to the normal state.
In the following, we argue that the structures, as shown in \fig{fig:pol}, are more likely to be observed experimentally compared to typically considered scenarios like LOFF state.

\section{Theoretical framework}
We consider 
Fermi gas on the spatial lattice with periodic boundary conditions at %
 finite temperatures. 
The stationary configuration has been found through
minimization of the grand canonical potential 
$\Omega = E - \mu_{\uparrow}N_{\uparrow} - \mu_{\downarrow}N_{\downarrow} - TS$,
where $N_{\uparrow\downarrow}=\langle \hat{N}_{\uparrow\downarrow} \rangle $,
and $\langle ...\rangle$ denotes the thermal average.
Entropy and temperature are denoted by $S$ and $T$, respectively.
The energy density functional has been chosen as (we set units $m=\hbar=k_B=1$):
\begin{align}
E = E_{\textrm{BdG}} = \int \left(
  \frac{\tau_{\uparrow}(\bm{r})}{2}
+ \frac{\tau_{\downarrow}(\bm{r})}{2}
+ g|\nu(\bm{r})|^2
\right)d\bm{r},
\end{align}
$\tau$ is defined as the kinetic energy density and $\nu$ is the anomalous density which can be associated with condensate 
wave function of Cooper pairs. These densities read ($\sigma=\{\uparrow,\downarrow\}$):
\begin{align}
\tau_{\sigma}(\textbf{r}) = \sum_{n} |\nabla v_{n\sigma}(\textbf{r})|^2 f(-E_n).
\end{align}
\begin{align}
\nu(\bm{r})=\dfrac{1}{2}\sum_{n} [u_{n\uparrow}(\bm{r})v_{n\downarrow}^{*}(\bm{r})-u_{n\downarrow}(\bm{r})v_{n\uparrow}^{*}(\bm{r})]f(-E_n).
\end{align}
For completeness we define normal density 
\begin{align}
n_{\sigma}(\textbf{r}) = \sum_{n} |v_{n\sigma}(\textbf{r})|^2 f(-E_n),
\end{align}
which is main experimental observable. The Fermi-Dirac function, $f(E) = \left(1+e^{E/T}\right)^{-1}$ is introduced to study finite temperature effects.
The single quasiparticle energies $E_n$ and wave functions $\varphi_n(\bm{r})=[u_{n\uparrow}(\bm{r}),u_{n\downarrow}(\bm{r}),v_{n\uparrow}(\bm{r}),v_{n\downarrow}(\bm{r})]^T$ are determined from the eigenequation
$\mathcal{H}\varphi_n(\bm{r})=E_n\varphi_n(\bm{r})$, where 
$\mathcal{H}$ is the Bogoliubov-de-Gennes (BdG) Hamiltonian:

\begin{align}\label{eq:bdg_ham}
\begin{gathered}
\mathcal{H} = 
\begin{pmatrix}
h_{\uparrow}(\textbf{r})-\mu_\uparrow & 0 & 0 & \Delta(\textbf{r}) \\
0 & h_{\downarrow}(\textbf{r})-\mu_\downarrow & -\Delta(\textbf{r}) & 0\\
0 & -\Delta^*(\textbf{r}) & -h_{\uparrow}^{*}(\textbf{r})+\mu_\uparrow & 0\\
\Delta^*(\textbf{r}) & 0 & 0 & -h_{\downarrow}^{*}(\textbf{r})+\mu_\downarrow
\end{pmatrix}.
\end{gathered}
\end{align}
In this case the mean-field hamiltonian is spin independent and reads $h_{\uparrow}(\textbf{r})=h_{\downarrow}(\textbf{r})=-\frac{1}{2}\nabla^2$ and $\mu_{\sigma}$ is the chemical potential for a given spin component ($\sigma=\uparrow,\downarrow$). 
The order parameter is related to the anomalous density $\Delta(\bm{r})=-g\nu(\bm{r})$, and $g$ is the s-wave coupling constant. 
In order to check stability of the predictions with respect to the choice of the energy density functional, we employ also more advanced form which is 
particularly suited for spin asymmetric systems at the unitary regime \cite{bulgac_2008,ufg_book,wlazlowski_2018,Kopycinski2021}.
This is so-called Asymmetric Superfluid Local Density Approximation (ASLDA) and reads
\begin{eqnarray}
E = E_{\textrm{ASLDA}} = \int  \Bigl( &\alpha_{\uparrow}(n_{\uparrow},n_{\downarrow})&
  \frac{\tau_{\uparrow}(\bm{r})}{2}
+ \alpha_{\downarrow}(n_{\uparrow},n_{\downarrow})
   \frac{\tau_{\downarrow}(\bm{r})}{2} + \nonumber \\
+  &D(n_{\uparrow},n_{\downarrow})& 
+ g(n_{\uparrow},n_{\downarrow})|\nu(\bm{r})|^2
 \Bigr) d\bm{r}.
\end{eqnarray}
As compared to the pure BdG approach, the above functional comprises
of polarization-dependent effective mass $\alpha_{\sigma}$,
the mean-field
term $D$, which scales with a density as a Fermi gas and is polarization dependent too. The paring coupling function $g$ also acquires density dependence. The functions $\alpha_{\sigma}$, $D$ and $g$ were chosen in such a way to assure compatibility with quantum Monte Carlo calculations for both spin-symmetric and spin-imbalanced unitary Fermi gas, for details see~\cite{ufg_book}.

Most of the results presented here were obtained within BdG
functional for 2D geometries. It was motivated from one side
by geometries of traps which can be nowadays realized experimentally~\cite{Kwon:2021a,arxiv.2204.06542},
but also by numerical complexity, which was reduced in this case to reasonable limits. In order to investigate the robustness of the results, we
have also performed selected 3D studies by applying full
ASLDA functional.
The coupling constant $g$ can be related to the scattering length. However, explicit relation is not needed in the context of these studies. Instead, we fix the value of $g$ in such a way to obtain $\Delta/\eF\approx 0.5$ when considering the spin-symmetric system ($\mathcal{P}=0$), which corresponds to the so-called unitary (strongly interacting) regime~\cite{PhysRevLett.128.100401}. The Fermi energy $\eF=\kF^2/2$ is related to the (2D) Fermi momentum $\kF=\sqrt{2\pi (n_{\uparrow}+n_{\downarrow})}$. The BdG equations were solved numerically on a spatial lattice of size $N_x\times N_y$ with lattice spacing $dx=1$ (definition of the length unit). The particle numbers $N_\uparrow+N_\downarrow$ were adjusted to satisfy $\kF\approx 1$. 
The minimization procedure starts with the uniform solution, which has been perturbed initially by the random
external potential $\Delta_{ext}(\textbf{r})=\Delta_{\text{r}}(\textbf{r})\cdot\exp\{i\phi_{\text{r}}(\textbf{r})\}$. The amplitude $\Delta_{\text{r}}(\textbf{r})$ has a positive real value constrained not to exceed $1\%$ of the pairing strength in symmetric system. Both, $\Delta_{\text{r}}$ and phase $\phi_{\text{r}}$ were randomly selected at each lattice point. The perturbation was switched off after 50 self-consistent iterations, which amounts to less than $5\%$ of the total simulation effort. Eventually, a converged solution was obtained in an unconstrained manner. The calculations were executed by means of the W-SLDA Toolkit~\cite{wlazlowski_2018,PhysRevLett.112.025301,WSLDAToolkit}.
In appendix~\ref{app:re} we provide reproducibility packs, allowing for numerical reproductions of the presented below data.

\section{Self-consistent solutions and their energies}
\begin{figure}[t]
	\includegraphics[width=0.5\linewidth]{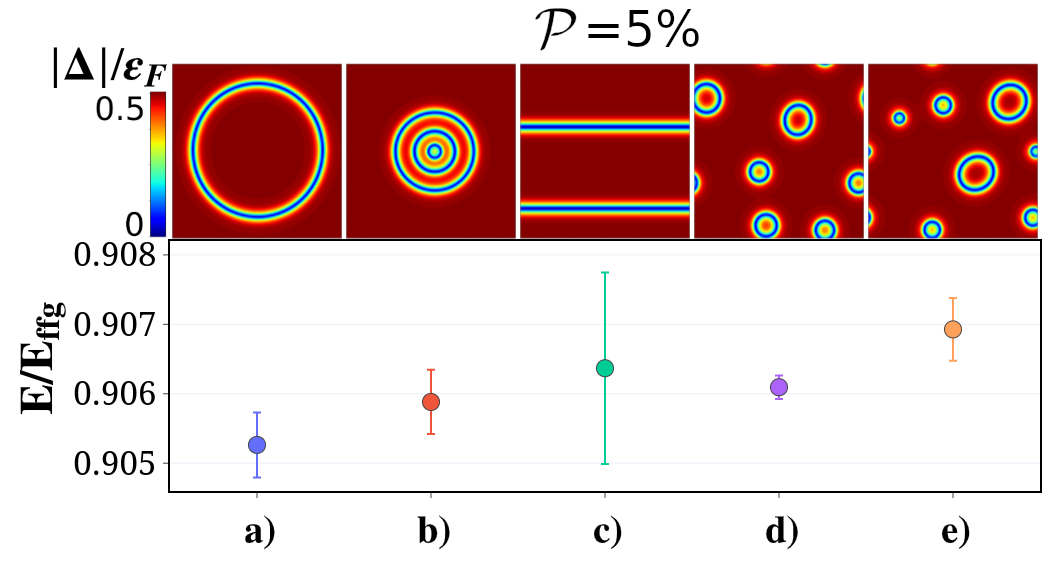}
	\includegraphics[width=0.5\linewidth]{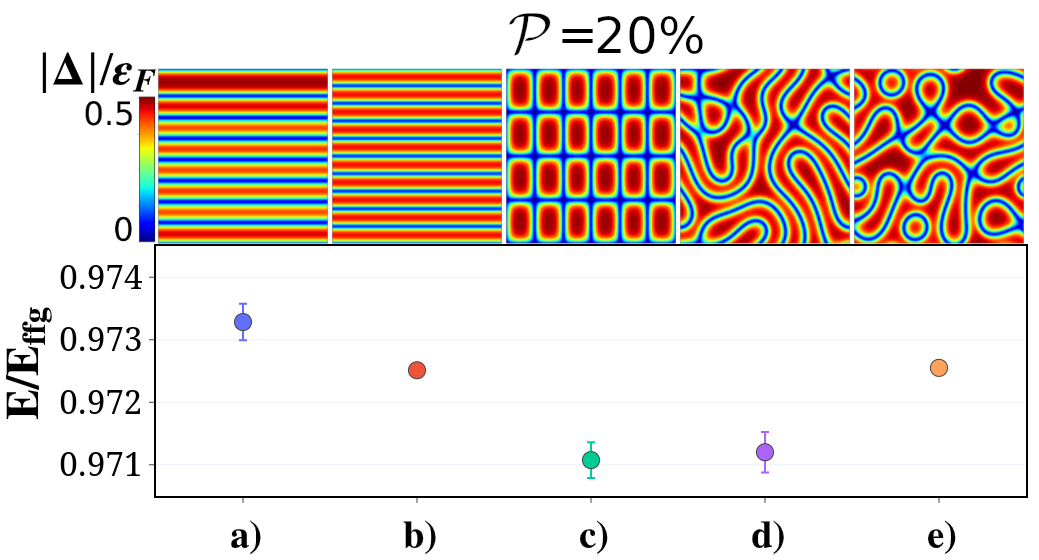}
	\caption{Spatial distribution of the order parameter (color maps) and corresponding energy for systems with $\mathcal{P}=5\%$ (top) and $20\%$ (bottom) population imbalances. All of the presented states represent self-consistent solutions of BdG equations obtained on a $100\kFi \times 100\kFi$ lattice with periodic boundary conditions. Cases a-c correspond to calculations where particular regular structure of $\Delta(\bm{r})$ 
was imposed. The cases d-e correspond to unconstrained calculations. The energies are scaled to the energy of non-interacting polarized Fermi gas $E_{\text{ffg}}=\frac{1}{2}N_{\uparrow}\varepsilon_{F\uparrow} + \frac{1}{2}N_{\downarrow}\varepsilon_{F\downarrow}$. Size of error bars 
denotes the numerical accuracy of energy determination in self-consistent calculations.
}
	\label{fig:P5}
\end{figure}

We have examined self-consistent solutions obtained according to two methodologies: using a random starting point (see above) and performing entirely unconstrained minimization or imposing a certain
constraint on the order parameter 
which allows us to obtain a self-consistent solution with required
symmetry.
Both approaches have limitations, and none guarantee that the lowest energy state is obtained. The unconstrained minimization process can get locked in a local minimum, while the constrained approach scans only a selected subspace of possible solutions. Examples of obtained solutions are presented in \fig{fig:P5} for two selected spin imbalances $\mathcal{P}=5\%$ (top) and $20\%$ (bottom) and with imposed constraints (a-c) or for the unconstrained process (d-e). The details on imprinting the nodal lines are presented in Appendix~\ref{sec:app:delta}.

In the regime of low spin imbalances (e.g., $\mathcal{P}=5\%$), the randomly seeded minimization process converges to states consisting of many ferrons randomly distributed in space, see~\fig{fig:P5} (d-e, top). Within this class of solutions, we find that states with a smaller number of ferrons (but larger in size) have lower energies. 
This can be understood as an additional energy cost related to the curvature of nodal lines favoring large ferrons~\cite{babaev}.
Guided by this observation, we have constrained the solver to search for a solution consisting of only one ferron in the simulation domain (a). We find that it has the lowest energy among all considered cases. Other identified cases, with comparable energy, are obtained if: the nodal lines preserve rotational symmetry, which can be linked to radial variant of LO state $\Delta(\textbf{r})\sim|\Delta(r)|\cos(qr)$ (b)~\cite{PhysRevA.103.053308}; the transitional symmetry is imposed, like in standard LO case (c). Note that the energy differences for these configurations are small. The ferron state has already been identified as the object of enhanced 
stability~\cite{ferron1}. Ginzburg-Landau approach, valid close to the critical temperature, also points to the state consisting of ferrons (called by authors {\it ring solitons}) as a candidate for the ground state~\cite{babaev}. Here we confirm that the state consisting of relatively large ferrons (radius by order of magnitude larger than the coherence length) is the predicted ground state in the low spin-polarization regime. 

In Ref.~\cite{ferron2}, it has been shown that in 2D, the ferron size has a linear dependence on the spin imbalance. Thus, as they grow with increasing the polarization, they will start to overlap, which we define as transition to medium polarization regime. Example of such case, $\mathcal{P}=20\%$, is shown in bottom panels (d-e) of \fig{fig:P5}.  
The obtained self-consistent solution depends on the initial random perturbation. They have comparable energies, which results in a vast possibility of different structures. We confront the energetics of disorder solutions with ordered ones. Here, the ordered structures are obtained by breaking the translational symmetry, e.g., imprinting a series of parallel nodal lines (a,b) or a nodal grid (c). Again, these ordered states can be interpreted as LO-like scenarios where the pairing field modulates as $\Delta(\textbf{r})\sim \cos{(\textbf{q}\cdot\textbf{r})}$ in one or both directions~\cite{bulgac_2008}. The main outcome of our calculations is that one cannot discriminate which one has lower energy: LO state or disorder state? Up to numerical precision, both of them have the same energies; compare panels c and d. This suggests that the ground state in this polarization regime is either highly degenerate or there exists a large number of states with energy almost equal to the ground state energy.   

\begin{figure}[t]
	\includegraphics[width=0.5\linewidth]{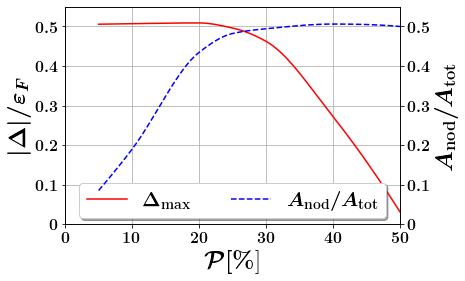}
	\caption{The maximal value of the pairing strength $\Delta_{\textrm{max}}=\max[|\Delta(\bm{r})|]$ (left axis), and the nodal area $A_{\textrm{nod}}/A_\textrm{tot}$ (right axis) as a function of the global polarization $\mathcal{P}$.}
	\label{fig:nodal_area}
\end{figure}

Nodal lines separate regions where the pairing field changes sign. Let us define nodal area $A_{\textrm{nod}}=\int \delta(\arg[\Delta(\bm{r})] - \pi)\,d\bm{r}$, so it measures an area where the phase of the pairing field is shifted by $\pi$. When expressing it as a fraction of the total area $A_\textrm{tot}=(N_x\times N_y)dx^2$ the relevant values span interval $[0-0.5]$. 
\fig{fig:nodal_area} shows that for $\mathcal{P}\gtrsim 30\%$, the fraction $A_{\textrm{nod}}/A_\textrm{tot}\approx 0.5$, which we define as the high polarization regime. In this regime the effective repulsion between nodal lines increases
and is directly related to the gradient term in Ginzburg-Landau model.
Therefore the system favors ordered lattice-like structures with equidistant
distribution of nodal lines.
Indeed the solutions of unconstrained calculations converge to structures 
where order parameter behaves like
$\sim\cos{(\textbf{q}\cdot\textbf{r})}$, at least locally, see \fig{fig:pol}.
It has to be noted that rapid cooling of highly polarized system may lead to domains
where the order is different due to Kibble--Zurek mechanism~\cite{zurek_1985}.
At the border of these domains the order parameter changes rapidly. 

With increasing polarization, the size of these ordered clusters increases and eventually becomes comparable with the system size. Not surprisingly, in the high polarization regime, the energy difference between LO state and states where LO order is present only locally is indistinguishable at the numerical level. Moreover, as the structures get denser, the average distance between two nodal points becomes comparable with the coherence length, with insufficient space for the pairing to recover fully within the structures. The maximal value of the pairing strength starts to decrease with the spin-polarization, eventually reaching a critical value, and the system transits to the normal state, see \fig{fig:nodal_area}.

\section{Robustness of the results}
\begin{figure}
	\includegraphics[width=0.5\linewidth]{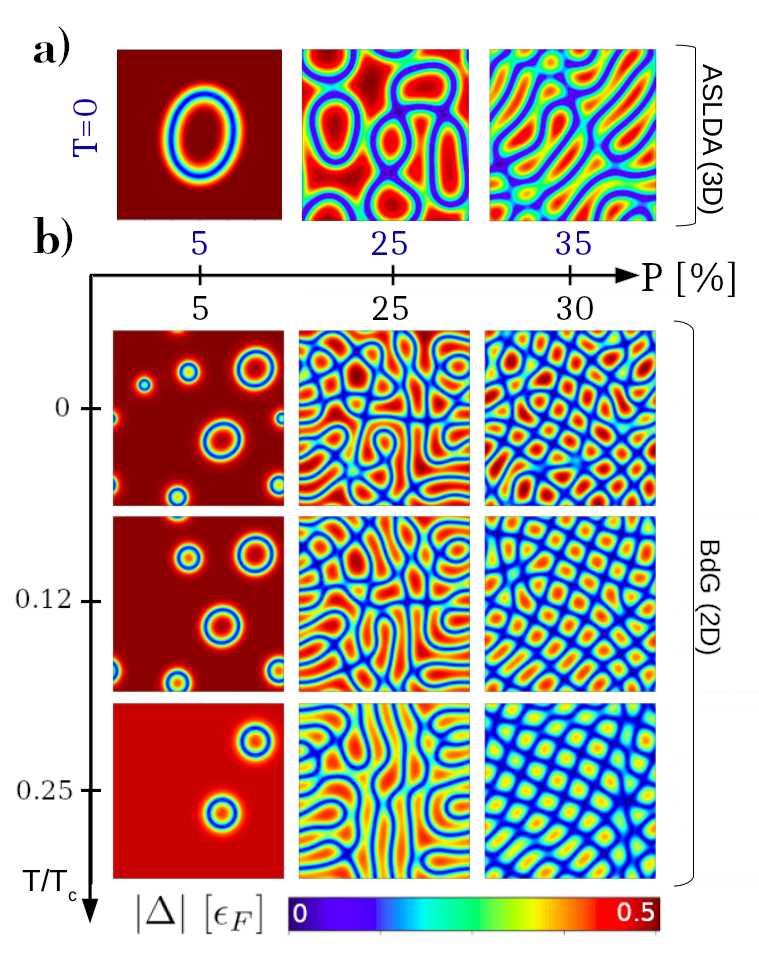}
	\caption{Example of disordered self-consistent solutions obtained by finite temperature BdG approach in 2D and 
ASLDA approach in 3D. The BdG eqs. we solved 
on a $100\kFi \times 100\kFi$ lattice, while ASLDA functional minimization was performed on a $60\kFi \times 60\kFi \times 16\kFi$ lattice with the system remaining translationally invariant along the third direction. Temperatures are expressed in units of critical temperature of superfluid-normal phase transition $T_c$ of uniform and spin symmetric gas.}
\label{fig:robust}
\end{figure}

Before we conclude, let us demonstrate the stability of the results. First, we checked the solutions' stability with respect to the temperature, see~\fig{fig:robust}. The state consisting of ferrons or disordered structures survive to temperatures
of the order of $T_{c}/4 - T_{c}/3$
and can compete energetically with the ordered solutions. As the temperature increases, the BCS coherence length increases as well, and structures whose sizes become comparable to this length scale melt. 
Except for this quantitative change, we do not find at finite temperatures any qualitative 
modifications as compared to the $T\rightarrow 0$ limit.    

Next, we confirmed that the similar structures are predicted if
we consider 3D unitary Fermi gas and apply more elaborate
energy density functional based on ASLDA. Imposing translational symmetry 
of the 3D system along one direction (by using ansatz $\varphi_n(\bm{r})\rightarrow \varphi_n(x,y)e^{ik_z z}$), we find that the disordered structures can be 
found as self-consistent solutions of ASLDA theory, see \fig{fig:robust}. The results of 3D calculations are represented with 2D cuts in the plane perpendicular to the $z$ axis. For example, a single ferron represented as a ring in 2D would correspond to a cylindrical tube in 3D. The quantitative 
differences between BdG and ASLDA approaches are related
to different values of $\mathcal{P}$, 
which separate low-medium-high polarization regimes. 

\begin{figure}[t]
	\centering
	\includegraphics[width=0.3\linewidth]{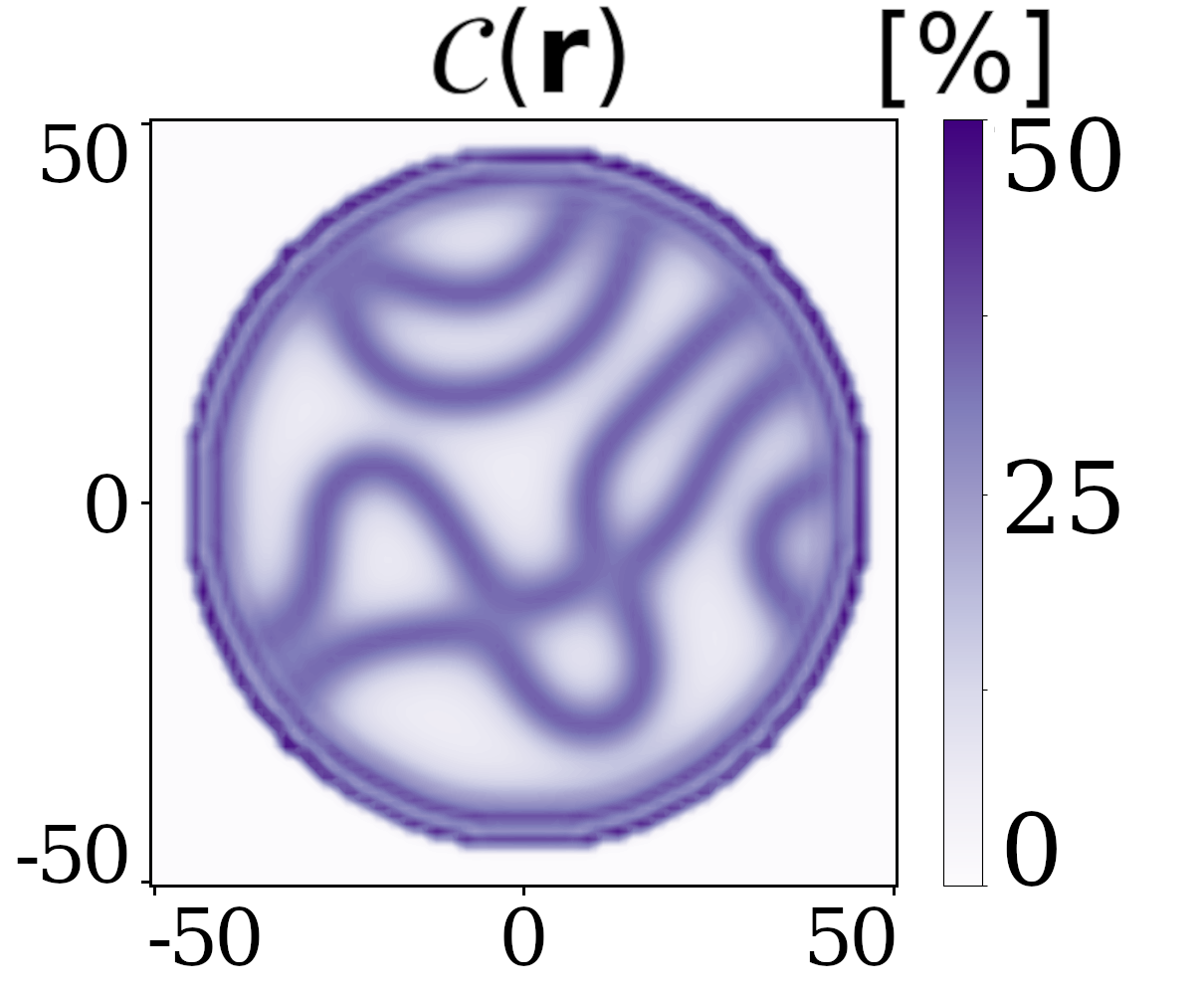}
	\includegraphics[width=0.3\linewidth]{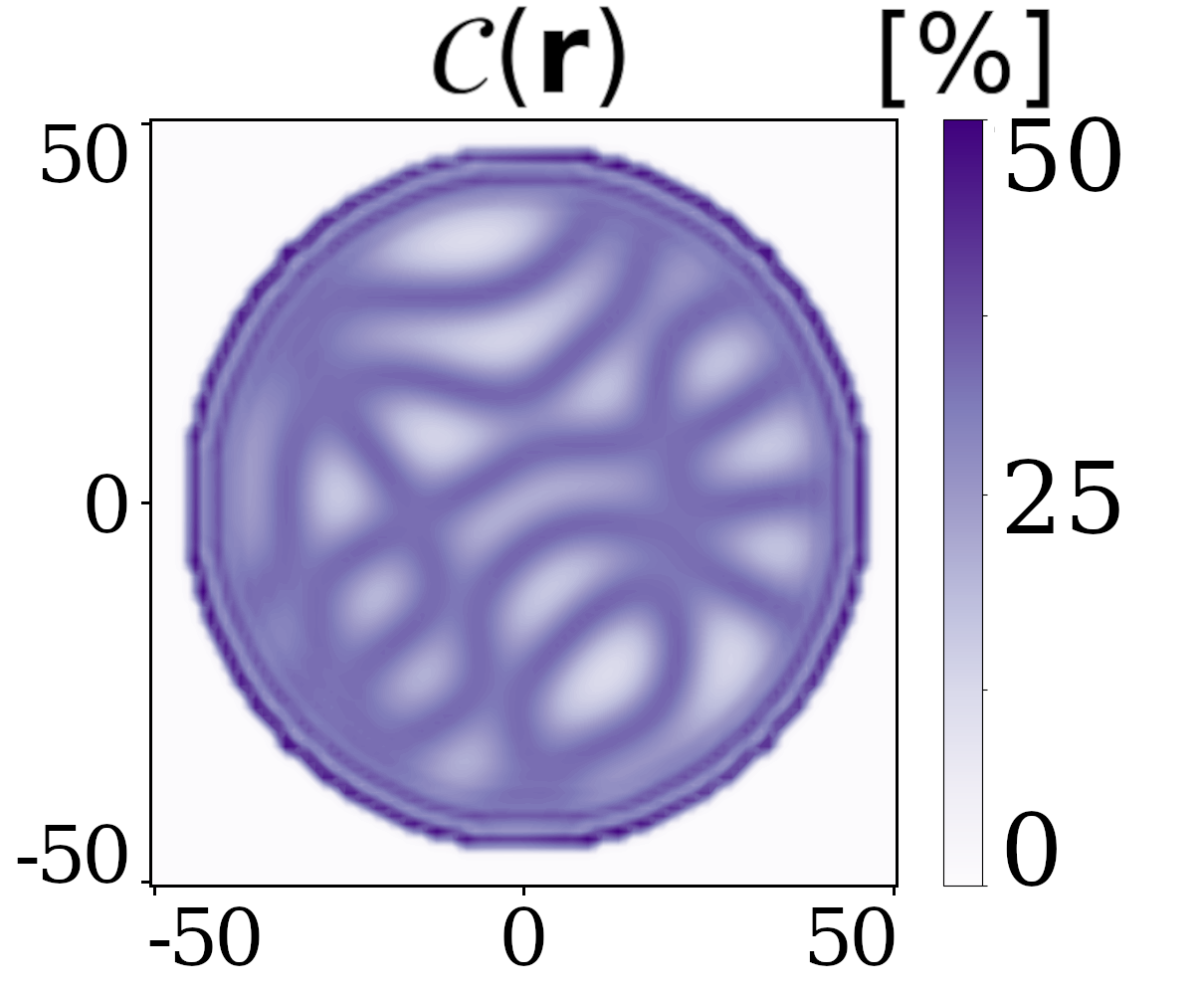}
	\caption{ Density contrast of a Fermi gas in a cylindrical box trap (of radius $46\kFi$) at $\mathcal{P}=25\%$ (left) and $\mathcal{P}=30\%$ (right). The self-consistent solutions are obtained by (2D) BdG equations, solved for $T/T_{c}=0.32$.}
	\label{app:fig:tube}
\end{figure}

Finally, we  
considered  
the box-like trap~\cite{Gaunt2013,Chomaz2015,Navon2016,Mukherjee2017,Navon_2021,Kwon:2021a}, to verify feasibility of an experimental detection. 
In~\fig{app:fig:tube}, we present the density contrast $C(\textbf{r})=[ n_{\uparrow}(\textbf{r})-n_{\downarrow}(\textbf{r}) ]/[ n_{\uparrow}^{(\mathrm{bulk})}+n_{\downarrow}^{(\mathrm{bulk})} ]$, where $n_{\sigma}^{(\mathrm{bulk})}$ is a density far away from the nodal line. For experimentally accessible temperature, of about $1/3$ of the critical temperature, we find that the contrast $C$ provides
strong enough signature to identify a disordered state, assuming 
imaging resolution to be of order of the coherence length. These estimates do not consider imaging integration along the third direction present in experimental realization, which may decrease the contrast. For sufficiently small heights of the trapping potential that is comparable to the coherence length (as in the experiment~\cite{Kwon:2021a}), the impact of averaging over the third dimension is expected to be negligible.

\section{Relaxation time scales}
The configurations that have been investigated do not necessarily form stable minima. It is likely that the plethora of configurations can evolve into one another, and the natural question arises concerning the time scale of this evolution. Moreover, as mentioned before, the order parameter in the high polarization regime may have a different structure within certain domains. This may appear due to fast cooling (Kibble--Zurek mechanism) creating rapidly varying order parameter at the boundaries between domains. These regions store additional energy, which can be released as the system evolves towards equilibrium.

\begin{figure}[t]
	\centering
	\includegraphics[width=0.5\linewidth]{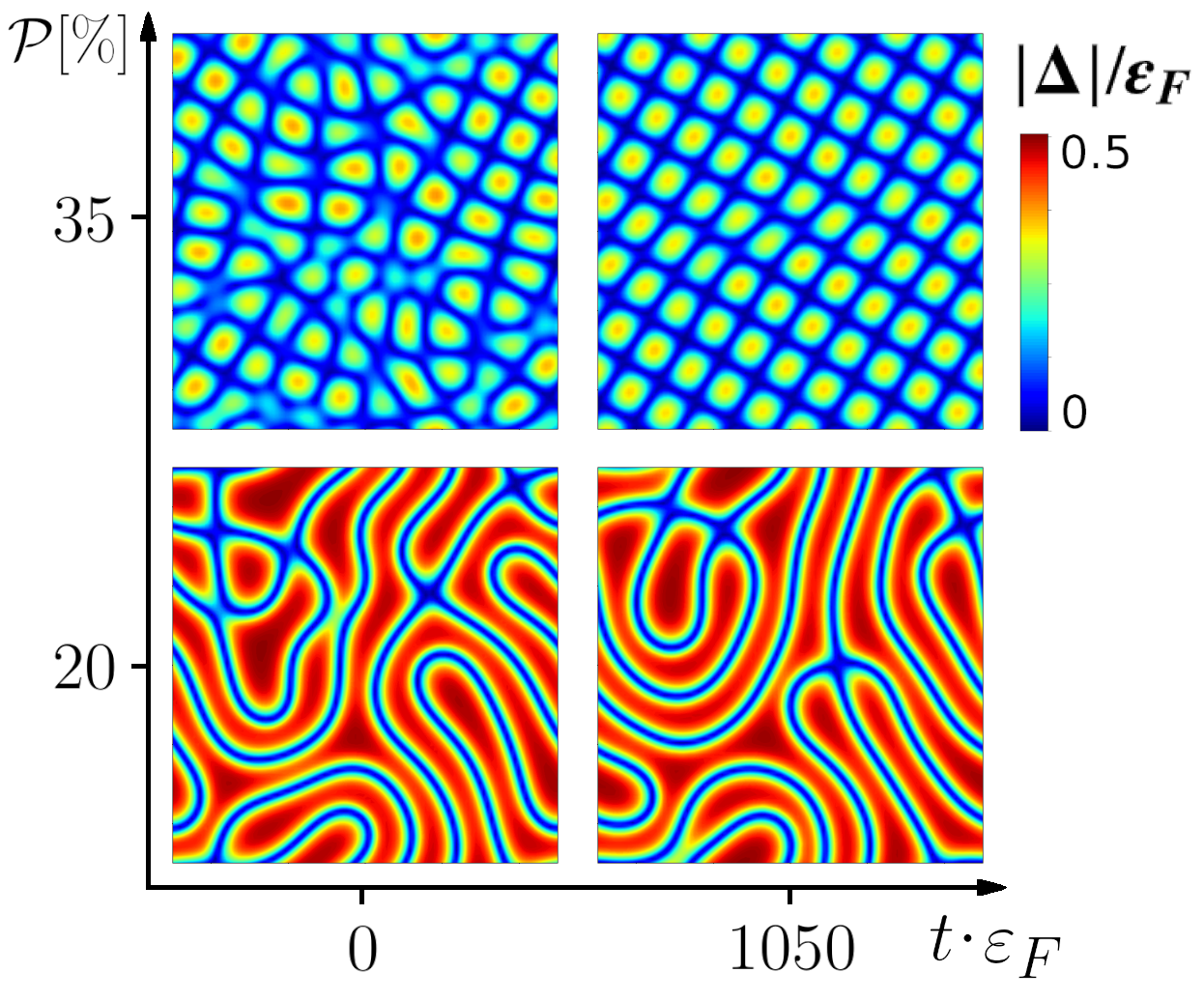}
	\caption{Two initial configurations (left panels) corresponding to $P=20\%$
	(bottom figures) and $P=35\%$ (top figures) evolved up to time $t\eF=1050$ on $100\kFi \times 100\kFi$ lattice.
	 The corresponding final configuration are shown in the right panels.}
	\label{td}
\end{figure}

In order to quantify these characteristic time scales, we have evolved the specific configurations corresponding to two regimes of medium and high polarization, as established in the previous sections. The results showing the change in the order parameter's structure within the evolution time $t\eF\approx 10^{3}$ are presented in ~\fig{td}. As expected, in the medium polarization regime, due to the weak interaction between nodal lines, we observe that the order parameter's structure is altered, but there is no qualitative change; the new configuration has a similar liquid crystal-like structure as the initial one. The situation is 
different for the high polarization regime. The initial system possessing discontinuity of the order parameter relaxes within the same time scale into a fully ordered system. Clearly, the strong repulsion between nodal lines leads to this effect. It should be noted that the final configuration in the high polarization regime shows a periodic modulation in the density and the order parameter with a global phase coherence which can be identified as a supersolid~\cite{PhysRevLett.25.1543,bulgac_2008}. 

\section{Conclusions}
We have demonstrated that the energy landscape of the spin-imbalanced system is very complex, with many 
configurations of comparable energies. We have explicitly demonstrated that the LOFF-type solutions represent only a tiny fraction of possible self-consistent solutions. The other identified solutions are characterized by disorder states, which cannot be parametrized in terms of simple ansatz for the pairing field, like $\Delta\sim \cos{(\bm{q}\cdot \bm{r})}$. 
The number of possibilities for arranging the nodal lines for disorder states is vast as compared to the ordered cases. During a cooling process, where we do not have control over the order parameter distribution, the system will likely settle in the minima corresponding to the disordered state. 
This scenario becomes even more likely when increasing the cooling rate. Then the Kibble-Zurek mechanism will naturally favor disordered structures. For these reasons, we expect 
lack of perfect ordering in experimentally produced spin-imbalanced
systems (especially in the low and medium polarization regime), contrary to 
standard predictions  
concerning the LOFF state detection, see~\cite{Kinnunen_2018} for a recent review.
Disordered 
structures are  
moreover not stable in a sense they will evolve in time.
In the low polarization limit the time evolution does not change qualitatively the order parameter distribution.
The situation
is different in the high polarization limit where the system evolves towards a globally regular lattice-like structure. All these issues challenge the experimental verification of the proposed scenario here. For example, averaging over many realizations will wash out the signal from the density distribution, as in each case, the system can relax towards different local minimums. 

Recent studies indicate that the LOFF phase is expected to be unstable against low energy fluctuations, like paring fluctuations~\cite{PhysRevB.97.134513,PhysRevLett.103.010404,PhysRevA.84.023611,PhysRevA.95.063626}. In these works, the LOFF state is defined as a state where all Cooper pairs have well-defined momentum $\pm \bm{q}$. Thus, our results are compatible with expected fragility of the LOFF, as we expect the disordered states, with no well-defined wave vector for the order parameter fluctuations, are the ones of experimental importance.

\begin{acknowledgments}
We thank G. Roati, M. Zaccanti, and M. M. Forbes for fruitful discussions. Calculations and data analysis were executed by BT and TZ. All authors contributed to research planning, interpretation of the results and manuscript writing.
This research was supported by PLGrid Infrastructure and LUMI Consortium. Part of calculations were executed at the Eagle supercomputer within grant No. 518 provided by Poznan Supercomputing and Networking Center (Poland) and on Piz Daint supercomputer based in Switzerland at Swiss National Supercomputing Centre (CSCS), PRACE allocation No. 2021240031. Numerical implementation was supported by IDUB-POB-FWEiTE-2 Project granted by the Warsaw University of Technology under the Program Excellence Initiative: Research University (ID-UB). This work was supported by the Polish National Science Center (NCN) under Contracts No. UMO-2017/27/B/ST2/02792 and UMO-2017/26/E/ST3/00428. BT acknowledges support from the Polish National Science Center (NCN) Grant No. 2019/34/E/ST2/00289.
\end{acknowledgments}

\appendix

\section{Nodal structures imprinting}\label{sec:app:delta}
\begin{figure}[!b]
\centering
\includegraphics[width=0.7\linewidth]{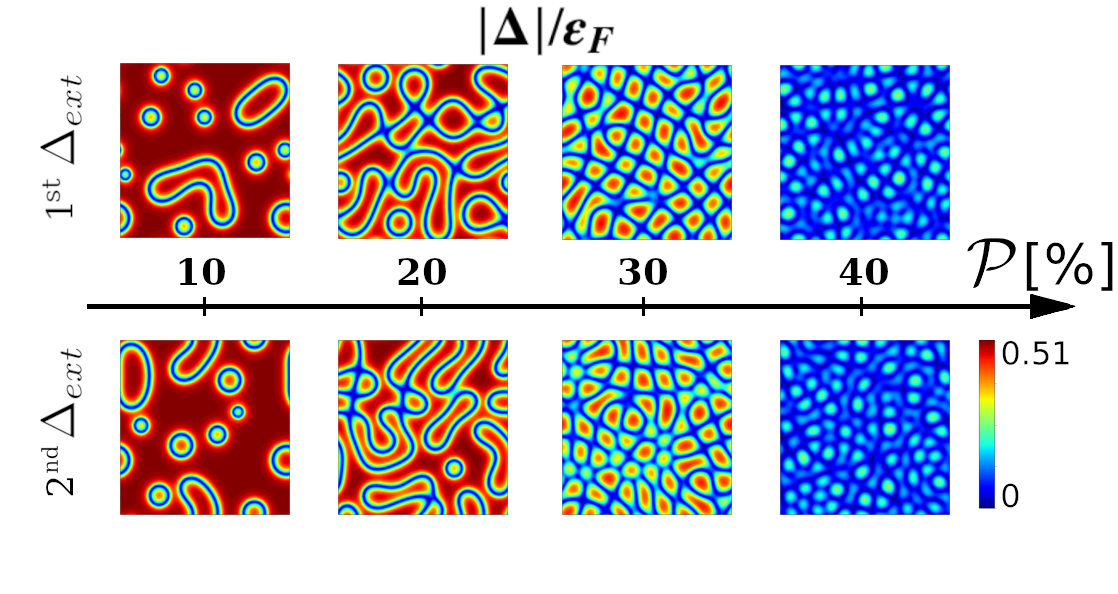}
\caption{Dependence of the final result concerning the choice of the random pairing field $\Delta_{ext}$. For each polarization, we observe qualitatively the same result. }
\label{fig:app:seeds}
\end{figure}

The self-consistent approach requires choosing the initial state, which in this study was the solution in the form of uniformly distributed gas. 
The uniform solution exhibits translational symmetry, which has to be broken. It is achieved with the introduction of an external component of the pairing field $\Delta_{ext}(\textbf{r})$. The total pairing field is then given by:
\begin{equation}\label{eq:delta_ext}
	\Delta (\textbf{r}) = -g_{\text{eff}}\nu (\textbf{r})
	+\Delta_{ext} (\textbf{r}) \theta (\tau - \text{it}).
\end{equation}
The external field  $\Delta_{ext}$ is chosen to be random
at each lattice point $\textbf{r}$ and fixed through $\tau=50$ self-consistent iterations, after which it is switched off. In the remaining iterations, the algorithm converges to a self-consistent solution without additional constraints. 
The amplitude $\max |\Delta_{ext}|$ is two orders of magnitude smaller than the pairing strength in the uniform system. 
We have repeated calculations with different noise choices to test the stability of the conclusions. 
The procedure produces qualitatively the same outcome, 
irrespectively of the shape of the initial perturbation, as shown in Fig.~\ref{fig:app:seeds}.

The random initial perturbation is used whenever it is desired to break all spatial symmetries. In selected cases, we study solutions that obey desired symmetry or belong to a specific class of solutions. An example is a solution in the form of a large ferron. Such solutions are obtained by enforcing the pairing field with a $\pi$-phase change. The equation~(\ref{eq:delta_impr}) demonstrates a method of guiding the solver toward a solution in the form of a large ferron. 
\begin{equation}\label{eq:delta_impr}
	\Delta (\textbf{r}) = 
	\begin{cases}
		-g_{\text{eff}}\nu (\textbf{r}) &\text{for}\ r < R_1, \\
		-g_{\text{eff}}\nu (\textbf{r})\cdot (-1)^{\theta (\tau - \text{it})}  &\text{for}\ r > R_2.\\
	\end{cases}
\end{equation}
The constraint above forces the paring field to have phase shifted by $\pi$ (far away from the center $r>R_2$) in comparison to the value close to the center ($r<R_1$).  
Note that the prescription for the $\Delta$ is incomplete as its form is undefined for $r\in[R_1; R_2]$. In this way, we let the solver decide the optimal radius of the ferron at a given spin-polarization. 
After $\tau$ iterations ($\text{it}$), the imprinting scheme is turned off, and the code finds the ground state solution. The value of $\tau$ differs for various imprinting procedures, but it is usually approximately ten times lower than the duration of the self-consistent process. The similar methodology was applied when searching for solutions with other types of regularities. 

\section{Energy corrections}
Our self-consistent algorithm is designed to converge to a solution with the requested number of particles $\Nreq_\sigma$ by independently adjusting the chemical potentials $\mu_\sigma$.
Practically, the constraint for the particle number is satisfied only with some accuracy
\begin{equation}\label{eq:CONV_npart}
		\forall_{\sigma}\quad \epsilon_{N, \sigma} = \frac{ \left\vert N_{\sigma} - \Nreq_{\sigma} \right\vert }{ \Nreq_{\uparrow}+\Nreq_{\downarrow} }.
\end{equation}
Although $\epsilon_{N, \sigma}$ is very small, the differences in desired and obtained numbers of particles should be considered when comparing energies between states of the same spin-polarization. Hence, we introduce a correction to the energy:
\begin{equation}
	\delta E = \sum_{\sigma=\updown} \mu_{\sigma} \left( \Nreq_{\sigma} -N_{\sigma}\right),\\\label{eq:deltaE}
\end{equation}
and uncertainty generated by this correction
\begin{equation}
	u(\delta E) = \sum_{\sigma=\updown} \mu_{\sigma} \Nreq_{\sigma}\cdot \epsilon_{N,\sigma}.\label{eq:udeltaE}
\end{equation}
This uncertainty sets the size of error bars presented in the figures in the main text.

\section{Simulated annealing}\label{app:sa}
\begin{figure}[t]
\centering
\includegraphics[width=0.7\linewidth]{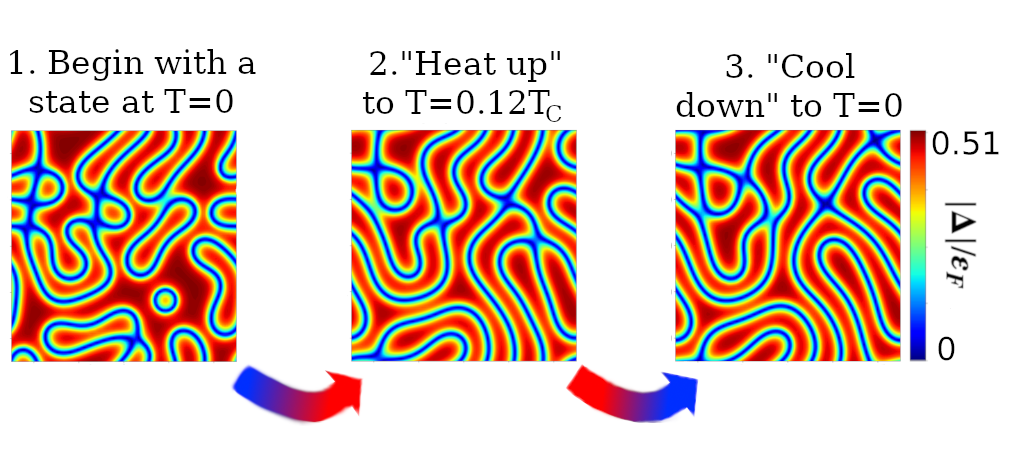}
\caption{Simulated annealing outline for the case $\mathcal{P}=20\%$ spin-polarization. Lattice size is $100\kFi\times 100\kFi$ with $k_F \approx 1$. } 
\label{fig:sa}
\end{figure}

To diminish the probability that the self-consistent solver terminates in a local energy minimum, we applied a technique similar to well known \textit{Simulated Annealing} (SA) method. 
It aims to take the system out of equilibrium, allowing for a significant position change in the energy landscape. The computation process is presented schematically in Fig.(\ref{fig:sa}). We executed calculations for $T=0$ until the algorithm found a self-consistent solution. Next, we increased the temperature to a new value and started the iteration process again. After a substantial number of self-consistent iterations, the temperature was brought back to zero. Then, the solver searched again for the ground state solution, starting from a modified finite-temperature state. The value of the finite temperature was chosen to make the smallest ferrons unstable, simultaneously leaving larger structures relatively unchanged. 

\section{Time evolution}
Dynamical simulations are done by employing the Time-Dependent Bogoliubov-de Gennes (TDBdG) framework:
\begin{align}\label{eq:bdg_ham_app}
\begin{gathered}
i \frac{\partial }{\partial t} 
\begin{pmatrix}
u_{n,\uparrow}(\textbf{r}, t)\\
v_{n,\downarrow}(\textbf{r}, t)\\
\end{pmatrix}
= 
\begin{pmatrix}
h_{\uparrow}(\textbf{r}, t)&  \Delta(\textbf{r}, t) \\
\Delta^*(\textbf{r}, t)  & -h_{\downarrow}^{*}(\textbf{r}, t) 
\end{pmatrix}
\begin{pmatrix}
u_{n,\uparrow}(\textbf{r}, t)\\
v_{n,\downarrow}(\textbf{r}, t)\\
\end{pmatrix}.
\end{gathered}
\end{align}
Single-particle hamiltonian $h$ and pairing field are defined according to formulas provided in the main text, with the main difference that now all densities also acquire time dependence. 
During the computation, the total energy and particle numbers' conservation laws were monitored. They were satisfied up to high precision of $2\cdot 10^{-7}$ for $N_{\sigma}(t)/N_{\sigma}(0)$ and $5\cdot 10^{-6}$ for $E(t)/E(0)$.
A detailed description of the numerical framework for TDBdG can be found in \cite{wlazlowski_2018}.

\section{Reproducibility packs} \label{app:re}
We include a set of self-explanatory examples for readers to be able to redo calculations presented in a publication on their own. We prepared supplementary files comprising all details required to reproduce three representative types of our calculations.
\begin{itemize}
\item \textit{BdG\_P25\_disordered.tar} - an example of a disordered state's creation with energy minimization according to the BdG energy density functional.
\item \textit{BdG\_P20\_grid.tar} - the one where grid-like pairing potential is enforced.
\item \textit{ASLDA\_P25.tar} - an example of 3D simulation with ASLDA functional.
\end{itemize}
Each of them requires access to considerable computing resources with W-SLDA toolkit \cite{PhysRevLett.112.025301, wlazlowski_2018, WSLDAToolkit} installed.

\end{document}